\documentstyle[12pt,worldsci]{article}
\font\tenbf=cmbx10
\font\tenrm=cmr10
\font\tenit=cmti10
\font\elevenbf=cmbx10 scaled\magstep 1
\font\elevenrm=cmr10 scaled\magstep 1
\font\elevenit=cmti10 scaled\magstep 1

\renewenvironment{thebibliography}[1]
 { \elevenrm
   \begin{list}{\arabic{enumi}.}
    {\usecounter{enumi} \setlength{\parsep}{0pt}
     \setlength{\itemsep}{3pt} \settowidth{\labelwidth}{#1.}
     \sloppy
    }}{\end{list}}
 
\begin{document}


\begin{center}{{\tenbf TOPOLOGICAL COMBINATORICS\\}
               \vglue 3pt
         {\tenbf OF A QUANTIZED STRING GRAVITATIONAL METRIC\\}  
\vglue 5pt
{\tenrm WAYNE R. LUNDBERG\\}
\baselineskip=13pt
{\tenit Physics Department, Wright State University,\\}
\baselineskip=12pt
{\tenit Dayton, Ohio 45435, USA\\}
\vglue 0.3cm
{\tenrm ABSTRACT}}
\end{center}
\vglue 0.3cm
{\rightskip=3pc
 \leftskip=3pc
 \tenrm\baselineskip=12pt
 \noindent
A string-theoretic structure of the standard model is defined having a 4-D 
quantum gravity metric consistent with topological and algebraic first 
principles.  Unique topological diagrams of string states, strong and weak 
interactions and quark families are evolved from this metric but published
separately.  The theoretical structure includes known static and dynamic 
symmetries.  A philosophical perspective on modern physics originates numerous 
opportunities for formal mathematical discussion.
\vglue 0.6cm}
\baselineskip=14pt
\elevenrm
	Consider a-priori the 
\underline{Theorem}: The topology of a three-point spinning metric 
is a necessary and sufficient basis for 4-dimensional physical objects.
The pregeometry of a two-point metric has been shown\cite{alv92} to provide the 
basis for physically interesting theories of quantum gravity.  Here $g_{\mu \nu}
= \langle g_{\mu \nu} (${Y}$_{r,y,b}) \rangle$.  Similar topological 
mechanisms of mass generation have been explored previously\cite{topo}.

	This topological metric is equated combinatorially with
Rishons\cite{har83} using two orthogonal intrinsic spin orientations.  Local 
colored gauge fields are identified without formal mathematical definition to 
be compatible with QCD.  Perturbations of the topological structure are not used
to construct the static and dynamic symmetries of Standard particles and 
interactions.

     Ten dimensional topological objects are constructed with the 3-point 
metric as a basis.  Three identical preons can be assembled in two distinct ways, 
corresponding to the electron (${\rm \bar T \bar T \bar T}$) and neutrino (VVV).
It is evident that the 10-D internal properties of the object literally ``curl 
up", ${\cal R}^{10} \supset {\cal R}^4$, to form the 3+1-D electron\cite{neem86}.
The formal
mathematical description of the topological electron is postulated to be rigid
\cite{fior_roberts}.  The string's ground state results from minimized internal 
tensions under a three-color partition, so it is called the Tripartite String 
Theory.  The usual\cite{gsw} Nambu-Goto action of the Tripartite String becomes 
$S= \mu \int_t \int_{color} \sqrt{h(}${Y}$_{r,y,b}) {\rm d}^2 \xi$.

	Rishon combinatorics are easily identified with the standard model 
group $SU(3)_c \times SU(2)_L \times U(1)_y$.  A material analog of this group 
is provided by restricting Rubic's cube.  Spontaneous symmetry breaking is 
analogous to the requirement that the corner cubies be restricted to their 
respective positions.  An electron is modeled by a left-hand twist of the r-y-b 
corner cubie, combinatorially similar to an isolated quark.  Charged quarks are 
modeled by twisting other cubies as follows.

The electric charge quantum numbers of particles are defined by the orientation
of the intrinsic curvature tensor.  Orientation may be graphically indicated by 
dark and light sided objects.  The quark's color is determined by the 
orientation of the spin axis with respect to the color gauge fields.  The 
oriented string allows two permutations of color order (r-y-b, r-b-y) which 
naturally models antimatter.

	The cubic symmetry of the electric and color charge diagram
is exploited to define the colored combinatorics of Rishonic string states.  A 
unique Rubic's cube color scheme is defined to map quark color and anticolor.  
Spin is modeled in this paradigm by rotation of the cube to avoid the 
superficial assignment of spin quantum numbers.

A Tripartite String proton diagram was copyrighted several years ago.  Color 
copies of the figures discussed in this paper may be requested via E-mail from 
lundberg@emsmtp.wpafb.af.mil.  The author assumes some artistic license by 
altering the scaling so that the Planck length is of the order of the proton 
scale.  The true scaling accounts for the relative success of open string 
theories\cite{lund}.

     The combinatoric model of a proton on Rubic's cube requires the operator 
$[\rm {(Y^-GYG^-)(G^-OGO^-)(O^-YOY^-)}]^2 $, where ${\rm G=\bar R}$.  Thus the 
algebraic group suggested by Golomb\cite{cube}
is defined.  Three such proton states exist which may be related by a strong
algebraic operation.  The spin content of the proton is evident even in the 
algebraic model\cite{close88}, although the majority of it stems from gluons.  
It is certainly a finite-sized constituent-quark model which is responsible for 
experimental measurements of spin dependance in proton collisions\cite{spin}.

	Consideration of the toroid-compact string and the spin-orientation 
basis of color charges determines the topology of the Tripartite String strong 
interaction.  This is a fundamental tree-level diagram 
wherein	the string is compacted to wind {\bf once} through the handle rather 
than around it.  The two-loop diagram is dissimilar to that of a 
second generation quark in that the world sheet self-intersects at the 
interaction point.  Triangulating the moduli space of closed-string geometric 
field theory is required at interaction vertices\cite{kaku}.

	An algebraic gluon operator such as $G_{y\to b}=\rm {B^-P_sO
(GB^-G^-B)^2O^-Y_sB}$ is applicable to the proton modeled previously.  Multiple 
gluon exchanges are modeled by superposition using the dynamical symmetry 
breaking rule that only r-y-b edge-cubie states are used.  Therefore only a 
defined subset of Rubic's cube (R-cube) is required for complete 
representation of $SU(3)_c \times SU(2)_L \times U(1)_y$.

	The toroid compact tripartite string allows three ``unmixed" knot 
states\cite{jones_w}.  The topological quantum gravity model does not allow 
world-sheet self-intersections for non-interacting strings.  The 
$\tau$ state resembles a trefoil knot and gives intuitive motivation to seek 
mass generation as a function of the {\bf volume} of the twisted toroid.  The 
three families explicitly evolved by the theory have been proven empirically.

	The six colored gluons are defined in the R-cube algebra.  All possible 
r-y-b edge states (gluons) have been mapped and shown to correlate identically 
with the three knot-states arrived at via the topological approach.  For a 
particle to be in a bound state a gluon must be present on the cube, disallowing
three colorless edge states.  The number of particulate states is now 256; 
fewer than the number of generators required for the standard model group.

	The 26-dimensional Tripartite String weak interaction world sheet 
comprises two 10-D closed, oriented, bosonic strings (4-D quarks) and a 
6-dimensional gauge superfield, or gluon\cite{boll89}.  Extrinsic curvature may 
be fixed while the intrinsic curvature tensor describes the shape of the quark,
which must change during weak interactions.  Thus temperature can be modeled by 
the extrinsic curvature of the world sheet\cite{ambj87}.

	A weak interaction may only be modeled on one R-cube with the 
stipulation that the two interacting particles be superimposed.  The two 
interacting quarks are operated on via an intermediate vector boson such as 
$\rm {[R_sP_s^-R^2](OR^-O^-R)^2}\times \rm {(RP^-R^-P)^2[R^2P_sR_s^-]}$.  
The assembly is then decomposed since no binding gluon exists and due to spatial
separation.  This method supplements that used in the Standard
Model with a more detailed notation of the operators used.

	The topological model of a $\pi^0$ meson reveals an `extra' symmetry.
The R-cube model also requires an extra symmetric operator to model 
$r \bar r$ quarks, i.e. $J_r={\rm G_sO^2G_s^2B_sY_s^2B_sG_s^2B^2R_s^2}$.  The 
total number of allowed states is now 512.  Since a colored supersymmetric 
operator is meaningless for the uncolored electron and neutrino, 16 cube states 
are disallowed, leaving the correct number of generators for the Standard Model.
\vglue 0.6cm
{\elevenbf \noindent 1. Conclusion \hfil}
\vglue 0.4cm
A description of a quantum gravitational string theory which evolves uniquely to
the Standard Model is given.  A mathematically well-founded proof of such a 
physical theory is potentially realizable in the near future.
\vglue 0.6cm
{\elevenbf\noindent 2. References \hfil}
\vglue 0.4cm

\end{document}